\documentclass[letterpaper, 10 pt, conference]{ieeeconf}
\IEEEoverridecommandlockouts
\usepackage[utf8]{inputenc}

\usepackage{algorithm}
\usepackage[noend]{algpseudocode}

\usepackage{pifont}
\newcommand{\cmark}{\ding{51}}%
\newcommand{\xmark}{\ding{55}}%

\usepackage{graphicx}
\usepackage{amsmath}
\usepackage{amssymb}
\usepackage{multirow}
\usepackage{cite}
\usepackage[caption=false,font=footnotesize]{subfig}

\DeclareMathOperator*{\argmin}{arg\,min}

\title{\LARGE \bf
High-Dimensional Controller Tuning through Latent Representations
}
\author{Alireza Sarmadi$^{1}$, Prashanth Krishnamurthy$^{1}$, and Farshad Khorrami$^{1}$
\thanks{$^{1}$The authors are with the Control/Robotics Research Laboratory (CRRL), Department of Electrical and Computer Engineering, NYU Tandon School of Engineering, Brooklyn NY, 11201. E-mail: \{as11986@nyu.edu, prashanth.krishnamurthy@nyu.edu, khorrami@nyu.edu\}}
}

\begin{document}

\maketitle

\begin{abstract}
In this paper, we propose a method to automatically and efficiently tune high-dimensional vectors of controller parameters. The proposed method first learns a mapping from the high-dimensional controller parameter space to a lower dimensional space using a machine learning-based algorithm. This mapping is then utilized in an actor-critic framework using Bayesian optimization (BO). The proposed approach is applicable to complex  systems (such as quadruped robots). In addition, the proposed approach also enables efficient generalization to different control tasks while also reducing the number of evaluations required while tuning the controller parameters. We evaluate our method on a legged locomotion application. We show the efficacy of the algorithm in tuning the high-dimensional controller parameters and also reducing the number of evaluations required for the tuning. Moreover, it is shown that the method is successful in generalizing to new tasks and is also transferable to other robot dynamics.
\end{abstract}

\section{Introduction}
\label{sec:intro}
Designing controllers for systems ranging across process control, automotive systems, and robotics is of great importance. However, due to nonlinear dynamics and internal and external perturbations, it can be difficult to find the desired controller for a wide range of tasks in complex systems (e.g., legged robots). Among the vast variety of control design methods in the literature, three main types of approaches can be distinguished. The first approach is the heuristic-based control in which an expert with a deep knowledge of the system designs the algorithm \cite{focchi2020heuristic,mac2016heuristic, huan2018adaptive, Bledt2020Heuristics} based on some heuristic. These algorithms suffer from the fact that it can be difficult to cover all the different cases that might be encountered during system operation.

The second approach is optimization-based control where the problem is formulated into an optimization problem. Model Predictive Control (MPC) is a well-known online framework for controlling various systems under uncertainties \cite{shim2003decentralized, tassa2012synthesis, borrelli2017predictive, biegler2021perspective, schwenzer2021review}. The performance of the MPC framework is very sensitive to cost function, constraints, and system dynamics. These hyperparameters should be tuned carefully to achieve the desired behavior. Usually, these hyperparameters are tuned manually or using grid search, which require many trial and error system evaluations and might not find the optimal set of hyperparameters. However, performing large numbers of trials on systems cause physical wear and tear and safety issues. One widely utilized solution to tune the hyperparameters with smaller numbers of system evaluations is to apply Bayesian Optimization (BO) \cite{shahriari2015taking}. BO is a black box optimization method to find the optimum of an unknown function when the function evaluation is expensive. BO has two main components: 1) a surrogate model for the objective function (usually Gaussian Process), and 2) an acquisition function that is used to decide where to pick the next sample. The main idea is to update the surrogate model when a new sample is obtained using Bayes' rule. Then, the next sample is picked using the acquisition function. One role of the acquisition function is to trade off exploration against exploitation. This makes BO a very interesting tool that can be used to adapt the policies designed in simulation with a few real-world experiments. However, BO's inference time grows as a cubic in the number of observations and exponentially with the dimension of the search space, therefore limiting its application in high-dimensional systems. Also, a high dimensional space results in an often heterogeneous function which makes the task of fitting a global surrogate model challenging. Due to the above reasons, BO is typically practical for a space with a dimensionality within around $10$ to $20$.

The third approach is learning-based control. Machine learning algorithms have become increasingly popular in recent years for designing control policies using data observed in an environment. More specifically, Deep Reinforcement Learning (DRL) \cite{arulkumaran2017deep} methods learn a controller without any prior knowledge about the system dynamics by training an end-to-end controller (i.e., the input to the controller is the observed state of the system, the output is the actuation vector). These learning-based methods typically need large numbers of samples (e.g., tens of thousands \cite{tan2018sim}) to achieve a reasonable performance. However, collecting large amounts of data on robots is challenging due to the physical wear and tear and safety issues in addition to the time required. Also, these algorithms can experience stability/performance degradations when adapting to new tasks \cite{khetarpal2020towards} and are very sensitive to hyperparameters \cite{henderson2018deep}.

In this work, we seek to improve the practical feasibility of the optimization-based control design approach through a learning-based framework to efficiently tune the high-dimensional controller parameter vectors. Our proposed method utilizes an off-line simulation environment to learn an underlying mapping of the high-dimensional controller parameter space to a lower dimensional space utilizing a Variational Auto Encoder (VAE) structure. This learned mapping then enables reducing the dimensionality of the search space for the tuning of the controller. To generate training data for learning the mapping, a sample-efficient BO-based search is utilized in an actor-critic framework. The encoder part of the trained VAE then acts as the desired mapping that enables reducing the dimensionality of the search space for the tuning of the controller on the system. With this learned mapping, the sample-efficient BO-based method is then applied in combination with the VAE decoder to tune the parameters on the system with a significantly smaller number of system evaluations compared to if the BO-based search was applied directly to the system in terms of the original high-dimensional controller parameter space. 

\begin{figure*}
\centering
\includegraphics[width=0.8\linewidth]{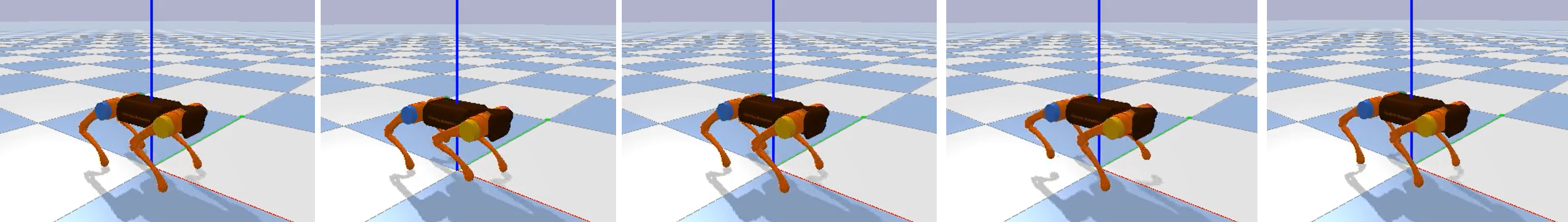}

\includegraphics[width=0.8\linewidth]{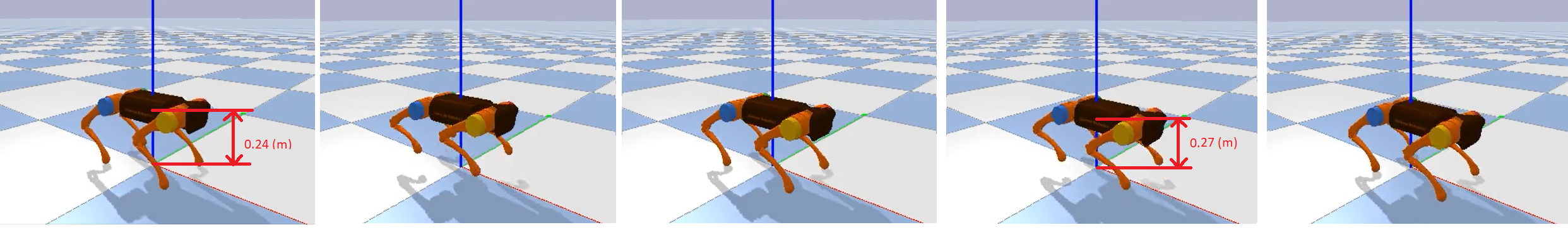}
\caption{Unitree A1 trotting (top row) and jumping (bottom row) motions in the PyBullet simulation environment. The maximum and minimum height of the robot during the jumping are shown.}
\label{fig:a1_trot}
\end{figure*}

\section{Related Works}
\label{sec:related_work}
Learning-based MPC has been studied recently~\cite{hewing2020learning} to improve its performance in presence of uncertainties in system model. Additionally, data-driven algorithms have been proposed to alleviate this problem by improving the system's model~\cite{aswani2013provably, soloperto2018learning}.
Another trend in learning-based MPC is to learn the design of the controller (e.g., cost function and constraints). \cite{marco2016automatic} proposed automatic LQR tuning using BO to tune the cost function parameters. In \cite{bansal2017goal}, the linear model of a system is learned using BO. In \cite{yeganegi2019robust, yeganegi2021robust}, the authors used BO to tune cost function parameters of the trajectory optimization problem. An Inverse Optimal Control (IOC) algorithm was proposed in \cite{rokonuzzaman2020learning} that learns hyperparameters of a defined cost function from human demonstration data for path tracking control problems. In \cite{marco2021robot}, the authors used BO for a jumping quadruped with motor current constraints.

Utilizing evolutionary algorithms \cite{jin2005evolutionary} is another approach for solving the optimization problem for high-dimensional spaces when the gradient of the cost function is not available (i.e., black-box function). However, these methods require a lot more samples. One of the fundamental work is Covariance Matrix Adaptation Evolutionary Strategy (CMA-ES) \cite{hansen2003reducing} that forms a parametric distribution over the solution space. Then, it chooses candidates to be evaluated by a black-box function from a parameterized search distribution. The first few best candidates are selected to update the parametric distribution.
Other papers in the literature use simulation to speed up BO. In \cite{rai2019using}, trajectories are generated during simulations to build a feature transformation.  Then, instead of passing the samples to the kernel function, they pass the transformed controller parameters. 

To implement the BO component in our work, we use Trust Region BO (TuRBO) \cite{eriksson2019scalable} instead of the standard BO that is only suitable for low-dimensional problems. This method uses a collection of simultaneous local optimization runs using independent probabilistic models to model the whole function by a combination of local surrogate models. These local surrogates allow for heterogeneous modeling of the objective function and do not suffer from over-exploration. To optimize globally, an implicit multi-armed bandit strategy is leveraged at each iteration to allocate samples between these local areas to determine which local optimization runs to continue.
 
\section{Method}
\label{sec:method}
We consider the general problem of tuning a vector of controller parameters corresponding to a parameterized controller for an agent interacting with an environment. 
For example, the set of parameters could include parameters in a cost function for an MPC controller. The objective is to minimize the cumulative cost of performing a control task.
For brevity, we will simply use the term ``parameters'' below to refer to the vector of controller parameters.
The agent interacts with the environment via the action $u_i\in A$ generated by the policy $\pi_{\theta}(x_i)$ in which $x_i\in S$, $\theta$ is the parameter vector, $S$ is the set of all states, and $A$ is the set of all the actions. The agent seeks to find the optimal $\theta$ by solving the optimization problem:
\begin{equation}
    \theta^*=\argmin_{\theta} L^{\pi_{\theta}} 
    \label{eq:theta_min}
\end{equation}
where the path cost of the policy (i.e., $L^{\pi_{\theta}}$) is defined as
\begin{equation}
    L^{\pi_{\theta}}(O^{des},O)=\sum_{i=0}^{N_H} q(O_i^{des},O)
    \label{eq:path_cost}
\end{equation}
with $N_H\in [0,+\infty)$ being the termination horizon of the problem, $q(.,.):S \times S \rightarrow \mathbb{R}$  the running cost, $O$ a task-specific vector of observed quantities from the environment that quantify performance, and $O_i^{des}$ the desired task behaviour in terms of the observed variables. For example, in the quadruped application, $O$ could be the velocity of the center of mass (CoM) while $O_{des}$ would be the desired value of the CoM velocity.

\begin{figure*}
	\centering
	\subfloat[]{\includegraphics[width=0.28\linewidth]{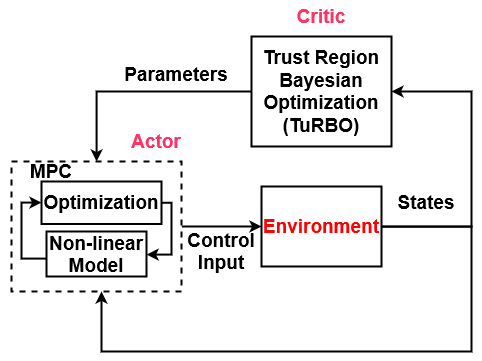}\label{a1}}
	\hspace*{0.5cm}
	\subfloat[]{\includegraphics[width=0.28\linewidth]{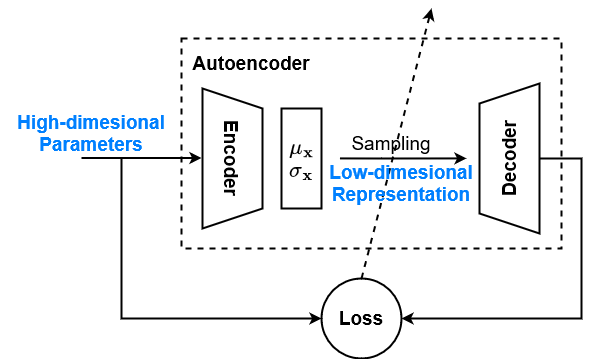}\label{a1}}
	\hspace*{0.5cm}
	\subfloat[]{\includegraphics[width=0.28\linewidth]{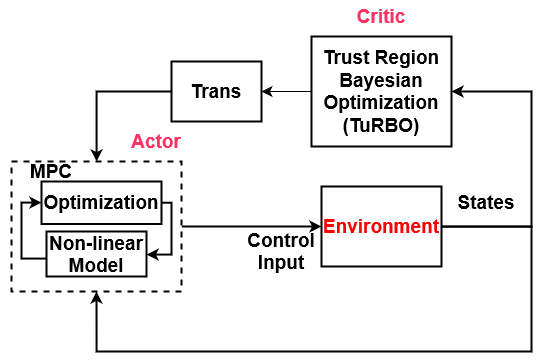}\label{a2}}
	\caption{Left: General Actor-Critic interaction in high-dimensional space; Middle: Training the autoencoder to learn a mapping from the high-dimensional parameter space to a lower-dimensional space; Right: Actor-Critic interaction in the lower-dimensional space.}
	\label{fig:ac_interact}
\end{figure*}

Our method approaches this problem in a gradient-free actor-critic manner in which the path cost is used to measure the performance of an MPC as an actor. A set of parameters is suggested by the critic. Then, the MPC actor generates the policy by solving Eq.~\ref{eq:mpc}. The interaction between actor and critic is depicted in Fig.~\ref{fig:ac_interact}. In the following, the critic and the actor are described in more detail.

\textbf{Critic:} The critic in our method evaluates the actor's performance using the path cost defined in Eq.~\ref{eq:path_cost}. The actor rolls out the policy and collects the (state, action) pairs to compute the path cost along the trajectory. Then, the critic suggests another $\theta$ in order to solve Eq.~\ref{eq:theta_min} after some iterations. Note that Eq.~\ref{eq:theta_min} can not be solved using gradient based methods, since we do not have access to the gradient of the $L^{\pi_{\theta}}$ w.r.t. to $\theta$. While a black-box optimization method such as BO would potentially be applicable, BO by itself is not suitable for high-dimensional parameter spaces as discussed in Sec.~\ref{sec:intro}. To overcome this limitation, TuRBO \cite{eriksson2019scalable} is utilized in which multiple independent local surrogate models represent the path cost instead of only one surrogate model. For each local model, a Gaussian Process (GP) is utilized to construct the local surrogate model
\begin{equation}
    L_l^{\pi_{\theta}} \sim GP_l(\mu_l(\theta), K_l(\theta,\theta))
\end{equation}
where $\mu_l(.)$ is the prior mean, and $K_l(.,.)$ corresponds to the kernel matrix ($K_l \in \mathbb{R}^{n \times n}$; where $n$ is the length of $\mu_l(.)$) of the $l^{th}$ GP model ($l\in \{1,\cdots,m\}$ where $m$ is the number of the regions, (i.e., the number of local models). The element at the $r^{th}$ row and $t^{th}$ column of the kernel matrix is the Matern kernel function \cite{melkumyan2011multi} between $\theta_r$ and $\theta_t$ defined as
\begin{equation}
    k_{Matren}(\theta_r,\theta_t)=\frac{2^{1-\nu}}{\Gamma(\nu)}(\sqrt{2\nu}d)\kappa_{\nu}(\sqrt{2\nu}d)
\end{equation}
where $d=(\theta_r-\theta_t)^T\Theta^{-2}(\theta_r-\theta_t)$ with length scale parameter $\Theta$, $\nu$ is a smoothness parameter, $\Gamma$ is the gamma function, and $\kappa_{\nu}$ is the modified Bessel function. The local surrogate models are optimized on a Trust Region (TR) centered at the best solution. The assumption here is that the local model can accurately model the function in the corresponding region. TuRBO considers hyperrectangles for the trust regions with an initial set of side lengths. During the optimization, these side lengths are rescaled based on the improvement. At each iteration, a set of candidates are chosen from the union of all the trust regions, then the $i^{th}$ candidate is chosen in a way that minimizes the function value across all the trust regions as 
\begin{equation}
    s_t = \argmin_s \argmin_{\theta\in TR_s} GP_s^t(\mu_s(\theta), k_s(\theta,\theta^{\prime}))
    \label{eq:turbo}
\end{equation}
where ${s_t}$ is the optimal trust region at iteration $t$, $\mu_s$ is the mean vector, $k_s$ is the kernel matrix for trust region $s$, and $GP_s^t$ is the GP model of the trust region $s$ at iteration $t$. The critic improves the GP model by obtaining more samples what are suggested by an acquisition function trading off the exploration vs. exploitation.

\textbf{Actor:} The actor is the trajectory optimizer that solves an optimal control problem in an MPC framework
\begin{equation}
\begin{aligned}
\pi_{\theta}= \min_{u_0,..,u_{N_1}} & J_T(x_{N_1}, u_{N_1}) + \sum_{i=0}^{N_1} J(x_k,u_k) \\
\text{s.t. } x_{k+1} &= f_d(x_k,u_k) \\
u_l                  &\le u_k \le u_u 
\end{aligned}
\label{eq:mpc}
\end{equation}
where $x_k$ and $u_k$ are state and control input at the $k^{th}$ time step, $J$ is the cost function, $J_T$ is the terminal cost function, $f_d(.,.)$ is the model of the dynamic system, $N_1$ is the MPC time horizon and $u_l$ and $u_u$ are the lower and upper bound vectors, respectively, for the inputs.

\subsection{Latent Representation}
If the aforementioned actor-critic framework is directly applied to the system with a high-dimensional parameters, a large number of system evaluations would be required. Moreover, for each new task, the framework needs to be repeated from scratch. To overcome these limitations, we learn a mapping from the high-dimensional controller parameter space to a lower-dimensional space in a data-driven manner. As in Section~\ref{sec:intro}, the intuition is that the well-performing parameters form a subset in a lower-dimensional space  can be learned in an unsupervised manner without assumptions on the mapping. For this purpose, the general actor-critic approach is applied and the set of controller parameters that provide reasonable performance (also referred to here as ``stable parameters'') is stored in a replay buffer called the training set. This set is utilized to learn the mapping by a VAE using a loss function of the structure
\begin{equation}
\begin{aligned}
    \min_{\psi, \phi} \frac{1}{N_t} \sum_{i=1}^{N_t} \| D(E(\theta_i;\phi);\psi) - \theta_i \|_2^2 + \\
    KL(N(\mu_{\overline\theta,\phi},\sigma_{\overline\theta,\phi}),N(0,I))
\end{aligned}
\end{equation}
where $E(.;\phi)$ is the encoder network with parameter vector $\phi$, $D(.;\psi)$ is the decoder network with parameters $\psi$, $N_t$ is the number of samples in the training set, $KL$ is the KL (Kullback–Leibler) divergence between a Gaussian distribution (with $\mu_{\overline\theta,\phi}$ as mean and $\sigma_{\overline\theta,\phi}$ as standard deviation) and a normal distribution $N(0,I)$. Here, $\mu_{\overline\theta,\phi}$ and $\sigma_{\overline\theta,\phi}$ denote the mean and standard deviation of the latent vector encodings (i.e., $E(\theta_i;\phi), i=1,\ldots,N_t$) of the training set $\overline\theta=\{\theta_i,i=1,\ldots,N_t\}$. 

The proposed method is described in Algorithm~\ref{alg:method} in which three procedures are defined. The $AC\_GEN$ uses $TuRBO$ to generate $N_{t}$ number of samples. This procedure takes the function to be optimized ($f$) as an input and the other input is the mapping function from the lower-dimensional space to the higher-dimensional space ($D$). This procedure first generates $N_{init}$ number of samples (which could include some manually tuned samples), then calls the $TuRBO$ algorithm to suggest new samples to be evaluated given previously seen samples. The $TRAIN\_VAE$ procedure trains a VAE with $D$ and $E$ as the decoder and the encoder neural networks, respectively. The parameters of the VAE (i.e., $D\circ E$ in which $E$ is the encoder neural network and $D$ is the decoder) are updated using Stochastic Gradient Descent (SGD) represented by the function $GU$ in the algorithm. The number of batches and the number of epochs are denoted by $n_{batches}$, and $n_{epochs}$, respectively. The $MAIN$ procedure corresponds to the overall execution of the algorithm in which $f$ is the path cost (i.e., the function defined in \eqref{eq:path_cost} for the application considered here) and the training samples ($(X_s, Y_s)$) are generated by $AC\_GEN$ called initially with no learned mapping (i.e., $Id$ which denotes ``Identity'' passed as the mapping function indicating that the parameter search is in the higher-dimensional space). Given the training samples, the decoder $D$ is trained by the $TRAIN\_VAE$ procedure. In the last phase, a new set of parameters is found ($X_r$) by taking advantage of the decoder. 

\begin{algorithm}
\caption{Proposed Method}
\label{alg:method}
\begin{algorithmic}[1]
\Require{BO~Cost}
\Ensure{$X_r, Y_r$}

\Procedure{AC\_Gen}{$f$, $D$}
    \State $X \leftarrow [x^i]_{i=1}^{N_{init}}$
    \State $Y \leftarrow f(D(X))$
    \For {$i \gets 1$ to $N_{t} - N_{init}$}
        \State $x_{new} \leftarrow TuRBO(X,Y)$
        \State $Append(X,x_{new})$
        \State $Append(Y,f(D(x_{new}))$
    \EndFor
    \State return X, Y
\EndProcedure

\Procedure{Train\_VAE}{$X$}
   \State Initialize D
   \Comment{Initialization of the Decoder Network $D$}
   \State Initialize E
   \Comment{Initialization of the Encoder Network $E$}
   \For {$i \gets 1$ to $n_{epochs}$}
       \For {$j \gets 1$ to $n_{batches}$}
           \State $GU(D\circ E, X^j)$ 
           \Comment{Updating the weights of the model $M=D\circ E$ using SGD}
       \EndFor
   \EndFor
   \State return D
\EndProcedure 

\Procedure{Main}{}
    \State $f \leftarrow BO~Cost$
    \State $X_s, Y_s \leftarrow AC\_GEN(f, Id)$
    \State $D \leftarrow TRAIN\_VAE(X_s)$
    \State $X_r, Y_r \leftarrow AC\_GEN(f, D)$
\EndProcedure

\end{algorithmic}
\end{algorithm}

\section{Results}
\label{sec:results}
\begin{figure}
\centering
	\subfloat[]{\includegraphics[width=0.5\linewidth]{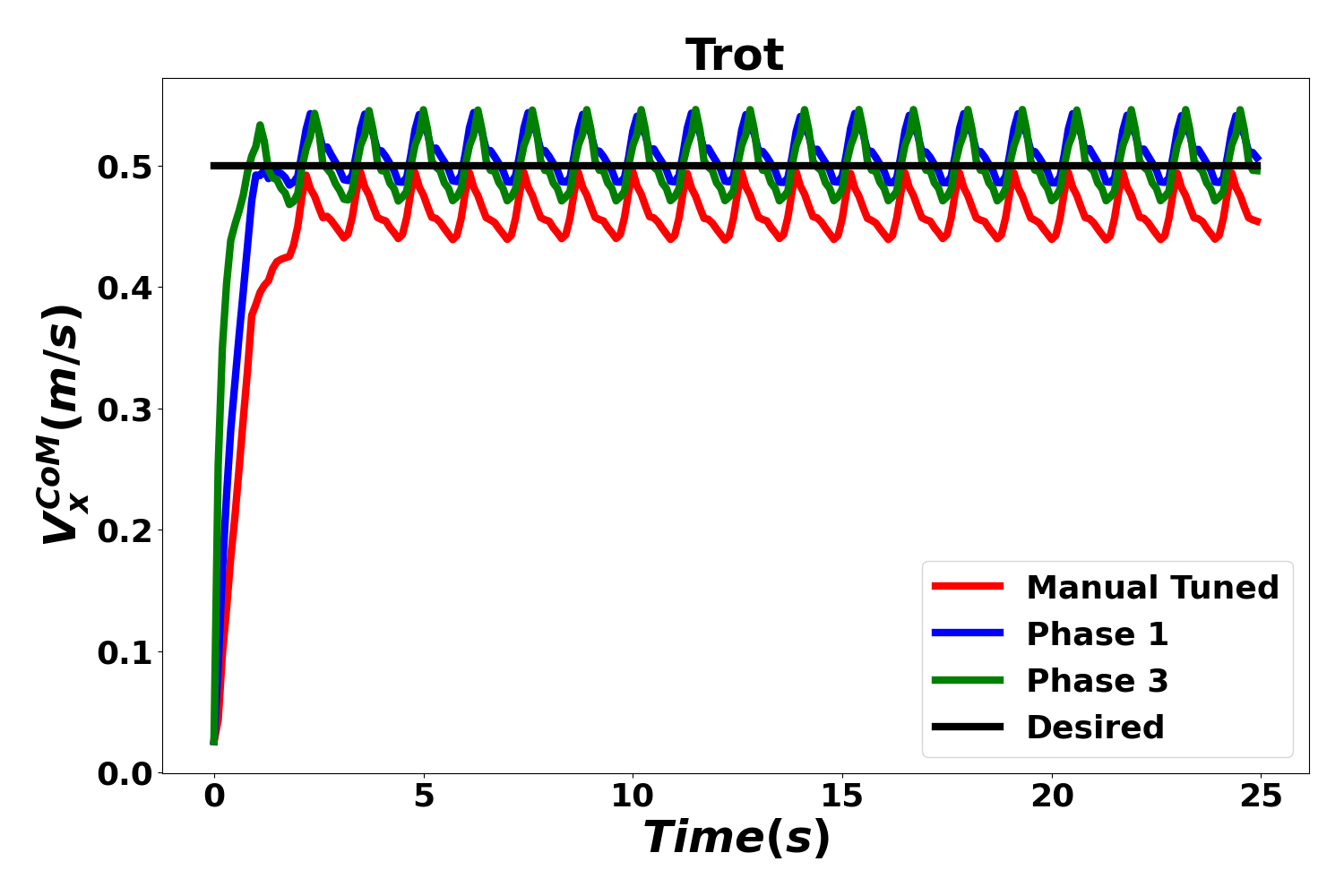}\label{a_trot}}
	\hfill
 	\subfloat[]{\includegraphics[width=0.5\linewidth]{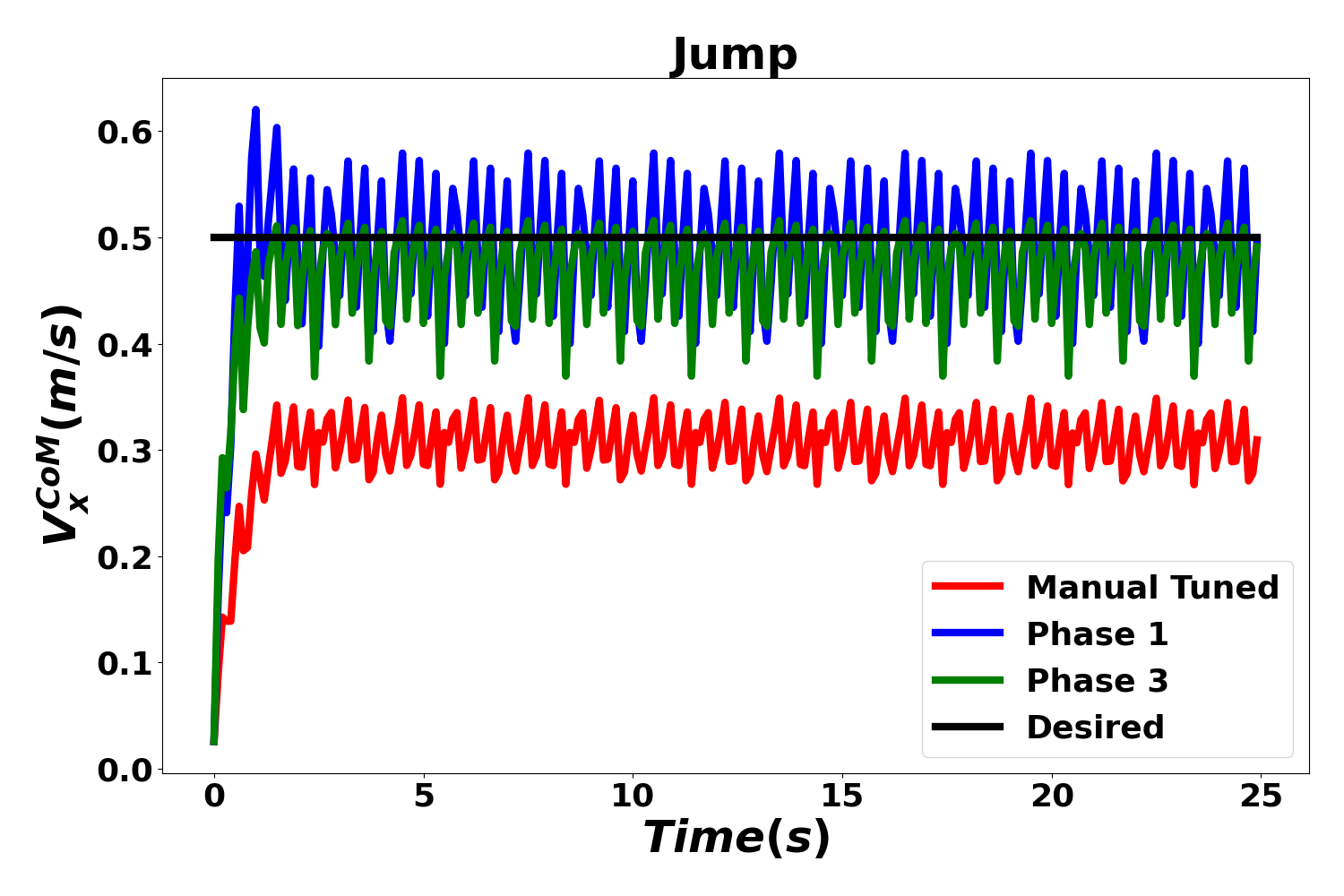}\label{a_jump}}
\caption{Comparison of Unitree A1 CoM velocity along x axis for manually tuned set of parameters after phases 1 and 3 for trot (top) and jump (bottom) motions. The desired velocity of CoM of the robot is 0.5 $m/s$.}
\label{fig:vcom_x_compare_trot}
\end{figure}

\begin{figure}
\centering
\includegraphics[width=0.6\linewidth]{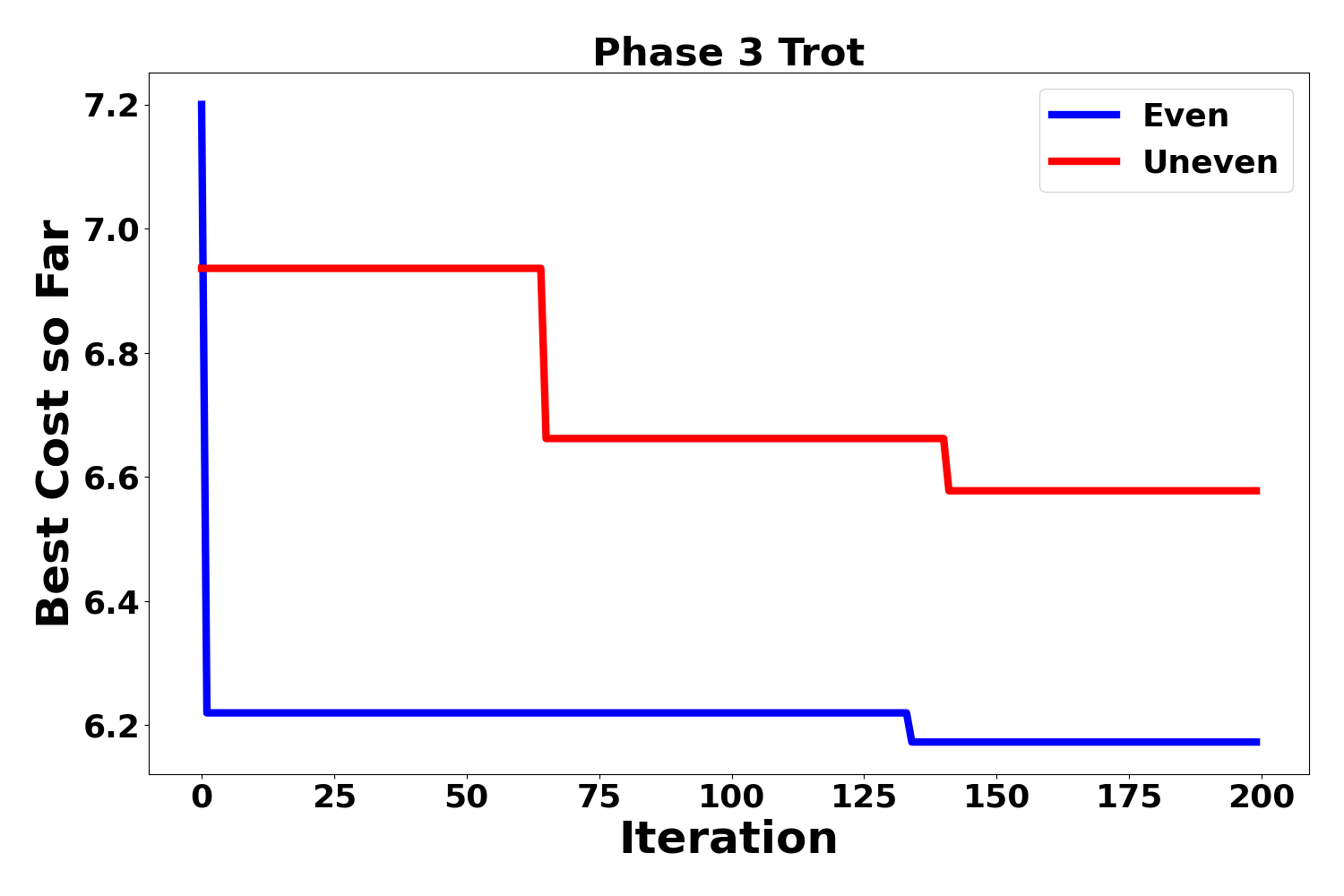}
\caption{Unitree A1's phase 3 best cost so far for 220 iterations of running the algorithm for trotting motion on an even and uneven surface. The final cost for the even surface is 6.17, and 6.57 for uneven surface.}
\label{fig:cost_turbo_3}
\end{figure}

We implement the proposed method on a quadruped robot for legged locomotion tasks. To show the efficacy of our proposed approach, we leverage two off-the-shelf MPC frameworks for legged locomotion: a state-of-the-art nonlinear MPC framework called BiConMP \cite{meduri2022biconmp} and the method proposed in \cite{di2018dynamic}. We have evaluated our method on the Unitree A1, B1, and the Solo12 \cite{grimminger2020open} quadruped robots in the PyBullet simulation environment \cite{coumans2016pybullet}. We consider the trotting and jumping motions (Fig.~\ref{fig:a1_trot}). In both motions, the robot moves forward with a desired speed of $0.5~m/s$ in the forward direction. For BiConMP on Solo12 robot, the controller parameter vector to be tuned is of length 77 while for Unitree A1 and B1, the controller parameter vector is of length 25. We  consider the running cost of the form
\begin{equation}
        q= 
\begin{cases}
    C_1,               & \text{if the robot falls}\\
    \| v_{CoM} - v_{Desired} \|^2,    & \text{otherwise}
\end{cases}
\end{equation}
where $C_1$ is a constant value, $v_{CoM}$ is the velocity vector of the CoM of the robot, and $v_{Desired}$ is the desired velocity vector that the user wants the robot to follow. It should be noted that falling is detected when the height of the CoM is outside the desired boundaries. We run 10000 iterations of the actor-critic method without dimensionality reduction (phase 1) for both trotting and jumping. The number of trust regions is set to 10. Once the parameter vectors are generated, those with a cost of less than 100 are chosen as training samples for VAE. The threshold 100 comes from the fact that the constant $C_1$ is chosen to be 100 (i.e., we pick parameters that do not at least result in falling). In this step, the chosen samples are used to train the VAE (phase 2). The architecture of the VAE is reported in Table \ref{table:ae_arch} in which $d_{high}$ and $d_{low}$ are the dimensions of the high-dimensional and low-dimensional parameter spaces, respectively. These dimensions are $d_{high}=77$ and $d_{low}=5$ for Solo12, and $d_{high}=25$ and $d_{low}=10$ for the Unitree robots. The VAE is trained with a 0.001 learning rate and 64 batch size. Mean Squared Error (MSE) loss is utilized as the loss function and optimized using Adam optimizer. Once the VAE is trained, its decoder is used to find the vector of controller parameters in a lower-dimensional space using the approach presented in Section~\ref{sec:method}. In the last phase of the algorithm, TuRBO is run for only 220 iterations with random initialization. In Fig.~\ref{fig:cost_turbo_3}, the accumulative minimum value is depicted for trotting motion. It can be seen that the algorithm successfully converges to a solution that has a relatively low-cost value. We removed initial unstable points from figures \ref{fig:cost_turbo_3} and \ref{fig:cost_turbo_3_generalization} for better visual comparison. Also, the velocity of the CoM of the robot along x axis is depicted in Fig.~\ref{fig:vcom_x_compare_trot}. It can also be seen that the set of parameters found in phases 1 and 3 are more successful in tracking the velocity compared to the manually tuned set of parameters.

\begin{table}
\begin{center}
\caption{The values $d_{high}$ and $d_{low}$ for the VAE architecture.
}
\label{table:ae_arch} 
\begin{tabular}{|c|cc|}
\hline
Model & Number of Nodes  &  Activation Function \\
\hline
\multirow{3}{*}{Encoder}  & $\lfloor \frac{d_{high}}{2} \rfloor$  & ReLU \\
 &  $\lfloor \frac{d_{high}}{4} \rfloor$ & ReLU  \\
 &  $d_{low}$ & Sigmoid \\
\hline
\multirow{3}{*}{Decoder} & $d_{low}$ & ReLU   \\
 &  $\lfloor \frac{d_{high}}{4} \rfloor$ & ReLU \\
 &  $\lfloor \frac{d_{high}}{2} \rfloor$ & Sigmoid \\
 \hline
\end{tabular}
\end{center}
\end{table}

\begin{table*}[h!]
    \centering
    \caption{Results for comparing the proposed algorithm with the state of the art algorithms.}
    \begin{tabular}{|c|c|c|c|c|c|}
    \hline
         \textbf{Method} & Max Dimension of the Parameters & \textbf{\#} of Experiments & Predefined Features & Transferability & Task Generalization\\
    \hline     
        \textbf{Our Method} & 77 & 75 & \xmark & \cmark & \cmark \\
    \hline     
        Yeganegi et al. \cite{yeganegi2021robust} & 4 & 50 & \xmark & \xmark & \xmark\\
    \hline
        Rai et al. \cite{rai2019using} & 50 & 100 & \cmark & \xmark & \xmark \\
    \hline     
        Rai et al \cite{rai2018bayesian} & 16 & 25 & \cmark & \xmark & \xmark \\
    \hline     
        Antonova et al. \cite{antonova2017deep} & 50 & 100 & \cmark & \xmark & \xmark \\
    \hline
    \end{tabular}
    \label{tab:comparative_study}
\end{table*}

\subsection{Real-World Simulation}
The proposed algorithm enables efficient parameter tuning on a quadruped by learning the mapping between the higher-dimensional parameter space and a lower-dimensional space. The last phase of the algorithm utilizes the learned mapping for performing the controller parameter tuning on the real system (i.e., to transfer the simulation-learned knowledge as encoded in the VAE to the real system). To evaluate the transferability robustness of the  algorithm from the simulation to the real world, external perturbations are added to the simulation environment. We considered uneven surface as a source of perturbation. A hilly surface (using Perlin noise with a maximum height of $0.02~m$) is added during phase 3 of the algorithm while in the first phase (which is based on the nominal simulation), there is no perturbation when generating the training set. Fig.~\ref{fig:cost_turbo_3} shows the accumulative minimum for 220 evaluations of the robot in the presence of surface perturbations while performing the trotting motion. Our approach is successful at finding an optimal solution even in the presence of the 
external perturbation. 

\subsection{Task Generalization}
\label{subsec:task_general}
\begin{figure}
\centering
\includegraphics[width=0.6\linewidth]{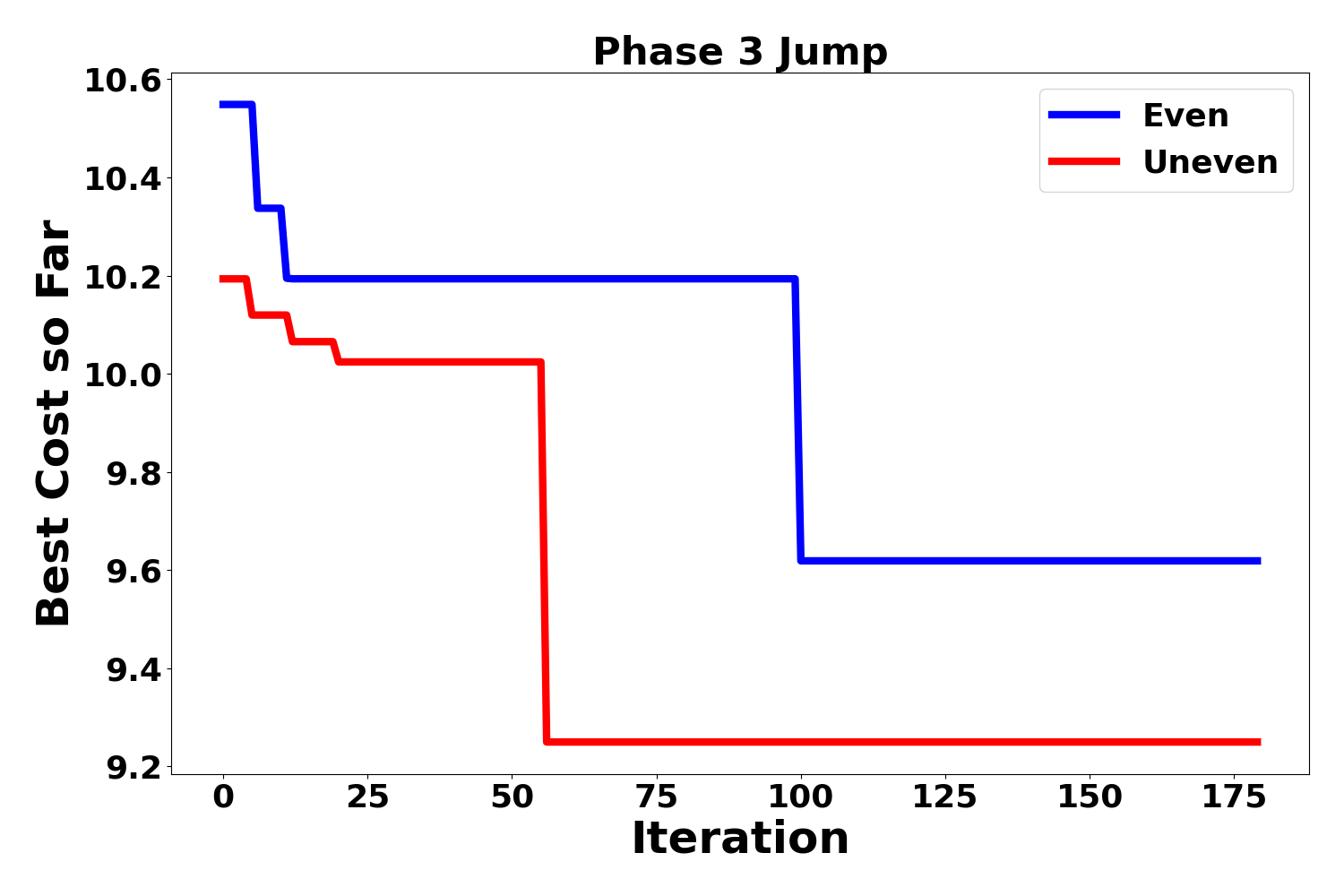}
\caption{Unitree A1's phase 3 best cost so far for 180 iterations of running the algorithm for jumping task using VAE trained on the trotting task. For even surface the final cost is 9.61, and for uneven surface is 9.25.}
\label{fig:cost_turbo_3_generalization}
\end{figure}

In this section, we refer to task generalization as utilizing simulation-learned knowledge (i.e., the learned mapping from high-dimensional controller parameter space to a lower-dimensional space) corresponding to one task to then tune the controller for a different task. For the legged locomotion task, the decoder is trained on a specific motion (e.g., trotting) and then used to find the parameters for another motion (e.g., jumping). The intuition in this task generalization study is that while the motion (i.e., specific control objective) is different, the reasonable subset of controller parameters will still be similar since it is the same underlying dynamics and therefore the learned mapping will still be somewhat valid even with the new motion (albeit requiring a bit more search during phase 3 tuning on the real system on the actual desired task).  We run phase 1 of the algorithm for trotting motion, then the VAE is trained in phase 2. Finally, the VAE is used for the third step of the proposed method to find the parameters for the jumping task. In Fig.~\ref{fig:cost_turbo_3_generalization}, the cumulative cost  is depicted for this case. It can be seen that the algorithm successfully found a parameter vector that results in a stable motion. Moreover, it is to be noted that when phase 1 of the algorithm is directly run for jumping motion, the error is improved after 2000 iterations while in this case, less than 200 iterations are needed. A comparison between different cases is reported in Tables~\ref{table:comparison_a1} and \ref{table:comparison_solo12} for Unitree A1 and Solo12, respectively. In these tables, the cost value (considering a  horizon of 5000 time steps in all results in this section) for a set of manually tuned parameters is reported in the second column. The best cost found in phases 1 and 3 and then phase 3 in presence of perturbation are given in third, fourth, and fifth columns, respectively.  The six column shows the best cost in phase 3 using a VAE trained on a different task in the sixth column.
It can be seen that for both motions, the minimum cost for both phases is less than the manually tuned set of parameters. It is seen that for all the cases except for task generalization, the tuned parameters using our approach results in less error than manually tuned ones. Furthermore, even for task generalization, our approach successfully yields parameters that achieve stable motion.

\begin{table}
\begin{center}
\caption{Path cost comparison for Unitree A1.}
\label{table:comparison_a1} 
\begin{tabular}{|c|c|c|c|c|c|}
\hline
\multirow{2}{*}{Method}& Manual &  Phase 1 & Phase 3 & Phase 3 & Phase 3\\
                       & Tuned  &          & (Even)  & (Uneven)& (Even)\\
\hline
Trot & 8.96 & 5.96 & 6.17 & 6.57 & 9.61 (Jump) \\
\hline
Jump & 13.81 & 8.21 & 9.52 & 9.15 & 20.57 (Trot)\\
\hline
\end{tabular}
\end{center}
\end{table}

\begin{table}
\begin{center}
\caption{Path cost comparison for Solo12.}
\label{table:comparison_solo12} 
\begin{tabular}{|c|c|c|c|c|c|}
\hline
\multirow{2}{*}{Method}& Manual &  Phase 1 & Phase 3 & Phase 3 & Phase 3\\
                       & Tuned  &          & (Even)  & (Uneven)& (Even)\\
\hline
Trot & 17.73 & 7.59 & 7.38  & 12.72 & 18.33 (Jump) \\
\hline
Jump & 12.24 & 8.41 & 10.26 & 10.88 & 10.17 (Trot)\\
\hline
\end{tabular}
\end{center}
\end{table}

\subsection{VAE's Mapping of Stable Regions}
\begin{figure}
\centering
\includegraphics[width=0.6\linewidth]{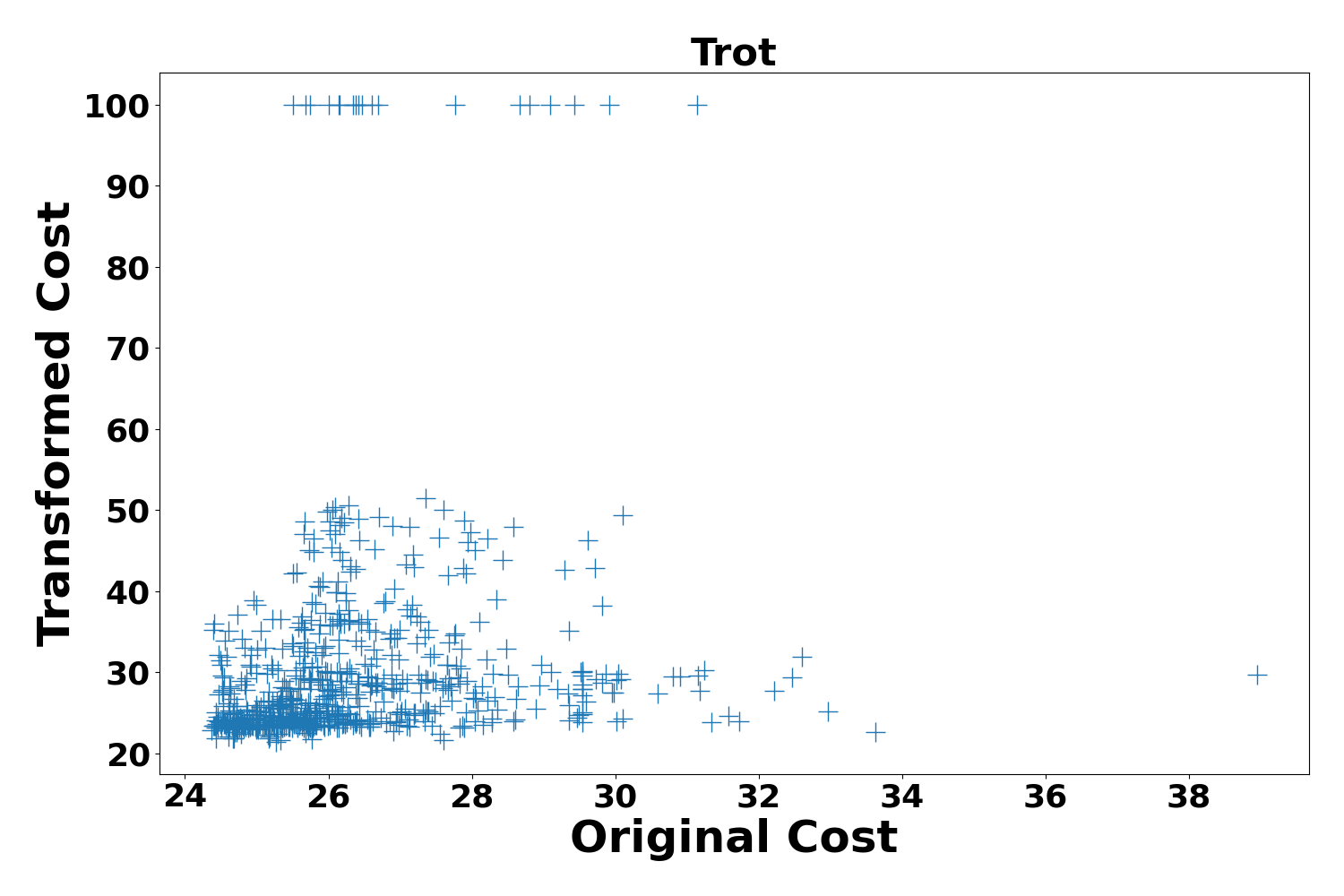}
\caption{The cost of transformed points found in phase 1 w.r.t. the cost of the same points is depicted.}
\label{fig:cost_point_compare}
\end{figure}

A crucial part of the proposed approach is the application of a VAE to find the informative latent representation of the controller parameters, especially over the ``good'' or ``stable'' subset of the parameter space in which reasonable performance is achieved. To specifically study this aspect of the VAE, we  run TuRBO to find a separate set of parameters that are not used for training the VAE. Then, we pass these parameters to the VAE to obtain the transformed set of parameters. The set of transformed parameters is evaluated on the robot to check if they result in a stable motion. Out of 677 stable points (i.e., parameter vectors that result in stable motion without falls) in the original space, 658 points are stable in the transformed domain (Fig.~\ref{fig:cost_point_compare}). It is therefore seen that over the stable part of the controller parameter space, the VAE achieves good reconstruction capability, therefore validating its use in focusing the search for controller parameters in the phase 3 of the proposed approach.

\subsection{Transferability to Other Robots}
\label{subsec:transferability}
In this section, we evaluate the transferability of the proposed algorithm. Specifically, we consider whether the mapping from high-dimensional to low-dimensional space learned for a specific robot can be potentially used for another robot. We train a VAE for the Unitree A1 robot while performing trotting. Then, this VAE is used for the Unitree B1 robot while performing jumping. The final cost is 12.98 after 220 iteration. The same procedure is done for finding parameters for the Unitree B1 robot while performing the trotting task. After 220 iterations, the final cost is 33.41. It should be noted that for this evaluation, we did not have access to a manually tuned set of parameters. This further illustrates the significance of the proposed method in that it can enable finding controller parameters for new robots on which even manual tuning has not been performed.

\subsection{Comparative Study}
We compare our proposed method with the state of the art algorithms reported in Table \ref{tab:comparative_study}. Unlike prior works, we do not consider predefined low-level features. We instead use a VAE to find a mapping from the high-dimensional controller parameter space to a lower-dimensional space using an approximate simulation model. Moreover, in our work, our mapping-based approach enables optimization of the actual high-dimensional parameter vector (e.g., a vector of length 77 for Solo12) while prior related works constrained the number of parameters to around a maximum of 50. Moreover, our method has the benefit of task generalization and transferability to other robots. These benefits comes from the fact that our method tunes the parameters in a low-dimensional space of extracted features. In comparison, other methods either need to define different low-level features \cite{antonova2017deep, rai2018bayesian, rai2019using}, or they need to solve the problem from scratch for different tasks or robots \cite{yeganegi2021robust}.

\section{Conclusion}
\label{sec:conclusion}
An efficient controller parameter tuning algorithm is proposed that is applicable to high-dimensional controller parameter spaces. The method is evaluated on legged locomotion tasks. The method is shown to be effective in finding parameters that result in stable motions with reduced numbers of system evaluations compared to direct tuning of the controller parameters. Also, it has been shown that the approach provides task generalization (transferring learned knowledge from a motion to find parameters for new motion tasks) and generalizability (transferring knowledge  learned on one robot to a different robot) properties. 

\bibliographystyle{IEEEtran}
\bibliography{references}

\end{document}